\documentclass{desyproc}

\begin{document}
\title{Neutrinos in Nuclear Physics}

\author{{\slshape R.~D.~McKeown}\\[1ex]
Jefferson Lab, Newport News, VA, USA\\
Department of Physics, College of William and Mary, Williamsburg, VA, USA }

\contribID{xy}

\confID{8648}  
\desyproc{DESY-PROC-2014-04}
\acronym{PANIC14} 
\doi  

\maketitle

\begin{abstract}
Since the discovery of nuclear beta decay, nuclear physicists have studied the weak interaction and the nature of neutrinos. Many recent and current experiments have been focused on the elucidation of neutrino oscillations and neutrino mass. The quest for the absolute value of neutrino mass continues with higher precision studies of the tritium beta decay spectrum near the endpoint. Neutrino oscillations are studied through measurements of reactor neutrinos as a function of baseline and energy. And experiments searching for neutrinoless double beta decay seek to discover violation of lepton number and establish the Majorana nature of neutrino masses.
\end{abstract}

\section{Introduction}
The discovery of neutrino oscillations in the distribution of atmospheric neutrinos by the SuperKamiokande experiment in 1998 \cite{SK_atmos} was a major event in the history of neutrino physics. This result established that neutrino flavors oscillate and that at least one neutrino type has a non-zero rest mass. Subsequently, further experimental studies of neutrino oscillations and masses were pursued with increased vigor and broader scope. Soon thereafter, the Sudbury Neutrino Observatory (SNO) reported the observation of solar neutrinos via the neutral current \cite{SNO}.  The SNO result showed that the total neutrino flux (summed over all three flavors) is consistent with expectation in the standard solar model, and that the $\nu_e$ flux is reduced due to flavor transformations, explaining the long-standing solar neutrino puzzle. Shortly after that, the KamLAND experiment reported the observation a deficit of reactor antineutrinos \cite{KamLAND} and subsequently a spectral distortion \cite{KamLAND2}, establishing that electron antineutrinos oscillate with a large mixing angle in a manner completely consistent with expectation based on the SNO results.

In the decade since these major discoveries, there has been a great deal of effort to develop a program of experiments to further explore the properties of neutrinos. The important remaining questions include:
\begin{itemize}
  \item What are the absolute values of neutrino masses (oscillation experiments only reveal squared mass differences $\Delta m^2$)?
  \item What is the correct ordering of the mass eigenstates ("normal" or "inverted" hierarchy)?
  \item Are the neutrino masses of a Majorana or Dirac type?
  \item What are the values of the mixing angles, and is there $CP$ violation in the neutrino mixing matrix?
\end{itemize}

\section{Absolute Neutrino Mass}

From neutrino oscillation experiments, we now know the values of $\Delta m^2$ \cite{PDG}:
\begin{eqnarray}
  \Delta m_{21}^2 &=& 7.5 \times 10^{-5} ~{\rm eV}^2 \\
  \Delta m_{31}^2 &=& 2.4 \times 10^{-3} ~{\rm eV}^2 .
\end{eqnarray}
Thus we can be sure that there is at least one neutrino mass eigenstate with a mass of at least $[\Delta m_{31}^2 ]^{1/2} \backsimeq 0.049$~eV. The endpoint energy in nuclear beta decay is modified by the effective neutrino mass
\begin{equation}
{m_{\nu_e}^{\rm (eff)}}^2 = \sum_i |U_{e i}|^2 m_{\nu_i}^2
\end{equation}
where the $U_{e i}$ are neutrino mixing matrix elements the sum is over all the experimentally unresolved
neutrino masses $m_{\nu_i}$. During the last decade experiments studying tritium beta decay have constrained this effective neutrino mass to be \cite{PDG}
\begin{equation}
m_{\nu_e}^{\rm (eff)} < 2~{\rm eV} .
\end{equation}
So there is presently a gap between 0.05~eV and 2~eV where experiments are needed to establish the absolute mass scale of neutrinos.

The distribution of matter in the universe depends sensitively on the neutrino contribution to the total matter density. Neutrinos are very light compared to all other particles, so at the epoch of structure formation they have a non-negligible thermal velocity, which controls their free-streaming length.  Since neutrinos do not clump on scales smaller than their free-streaming length this leads to smearing out of over-dense regions (structure) at small scales, leaving a characteristic imprint in the matter distribution. Current and upcoming surveys that probe the matter distribution can indirectly constrain or measure the sum of the neutrino masses. Current analyses constrain the sum of neutrino masses to be   $ \sum_i m_i < 0.23 {\rm eV} (95\% {\rm CL})$ \cite{CMB}. In the next decade there are good prospects to reach, via multiple probes, a sensitivity at the level of   $\sum_i m_i < 0.01 {\rm eV}$ \cite{CMB2} . Nevertheless, it is essential to address the neutrino mass scale below 2~eV in terrestrial experiments.

The KATRIN experiment \cite{KATRIN} is under construction at Forschungszentrum Karlsruhe and will provide measurements of the tritium endpoint spectrum with greater precision in the near future. This ambitious experiment utilizes a gaseous molecular tritium source. a pre-spectrometer to filter out lower energy electrons ($<E_0 - 0.3$~eV), a main spectrometer ( resolution 0.93~eV), and a position sensitive detection system. The apparatus is 70 meters long and the main spectrometer has a diameter of 9.8 meters. The experiment will be sensitive to neutrino masses $m_{\nu_e}^{\rm (eff)} > 0.2~{\rm eV}$ with 90\% CL (3 years running), extending the range of present knowledge by about an order of magnitude. Commissioning of the experiment is underway and KATRIN is expected to begin acquiring tritium decay data in 2016.

Improvements to the KATRIN experiment may be possible (for example using time of flight techniques) to further increase the sensitivity. However, the tritium source has reached the maximum density for transmission of the the 18~keV electrons of interest and a more sensitive spectrometer would need to be much larger than the main spectrometer of KATRIN. Therefore, it appears that another method may be necessary to make significant progress below 200~meV. A novel technique to detect the cyclotron radiation from a single electron in a uniform magnetic field using high-sensitivity microwave antennae has been proposed \cite{Project8}. Preliminary R\&D on this technique is in progress, and the first detection of cyclotron radiation from a single 30~keV electron has been reported at this conference and in \cite{Project8_2}.

\section{Reactor Neutrinos}
The neutrino mixing matrix contains 4 parameters: 3 mixing angles ($\theta_{12}$, $\theta_{23}$, and $\theta_{13}$) and a $CP$ violating phase $\delta_{CP}$. The combination of solar neutrino experiments and KamLAND have provided a value of $\sin_2 \theta_{12} \backsimeq 0.31$ \cite{PDG}. In addition, accelerator based long baseline neutrino experiments determine $\sin^2 \theta_{23} \backsimeq 0.39$ \cite{PDG}. While there is room for improvement in these determinations, much attention has been focused in recent years on measuring the remaining angle $\theta_{13}$. This problem has been effectively attacked by three reactor neutrino experiments: Double CHOOZ \cite{DC12}, RENO \cite{RENO12}, and Daya Bay \cite{DB}.

The formula for survival of electron neutrinos (or antineutrinos) in
the 3 flavor case is given by 
\begin{eqnarray}
\nonumber P(\nu_{e} \rightarrow \nu_{e}) &=& 1 - \sin^2 2 \theta_{13}(\cos^2 \theta_{12}
\sin^2 \Delta_{31} + \sin^2 \theta_{12} \sin^2 \Delta_{32}) \\
&\ & \ \ \ \ \ \ \ \ \ \ \ \ \ \ \  - \cos^4 \theta_{13} \sin^2 2 \theta_{12} \sin^2
\Delta_{12}
  \label{eq:surv3}
\end{eqnarray}
where $\Delta_{ij} \equiv \Delta m_{ij}^2 L / 4 E_\nu$.  Note that
the 2 terms  oscillate with different ``frequencies'' depending on
the values of the $\Delta m_{ij}^2$. Thus one can choose the
baseline $L$ to maximize (or minimize) the sensitivity to particular
$\Delta m_{ij}^2$. For an average reactor antineutrino energy of
4~MeV and a value of $\Delta m_{32}^2 = (2.43 \pm 0.13) \times
10^{-3}$~eV${}^2$ one finds that the optimum distance for the first
minimum is $L \simeq 2000$~m.

While all three reactor experiments have reported consistent values of $\theta_{13}$, the results from the Daya Bay experiment in China are the most precise. The Daya Bay nuclear power plant consists of 6 reactor cores in two groups (Daya Bay and Ling Ao) with a total thermal power capacity of 17.6~GW. The experiment includes 8 antineutrino detectors, each with 20~Tons of Gd-loaded liquid scintillator. Two detectors are located near (364~m) the 2 Daya Bay reactors and two are located near ($\sim 500$~m) the 4 Ling Ao reactors. Four detectors are located in the far experimental hall at 1912~m from the Daya Bay cores and 1540~m from the Ling Ao cores. The near detectors monitor the antineutrino fluxes from the two reactor groups so that the far detectors are sensitive to the degree of neutrino oscillations at the longer baseline. This method enables measurement of the oscillation effect with only slight sensitivity to the absolute flux of antineutrinos.

\begin{figure}[ht]
\centerline{\includegraphics[width=\textwidth]{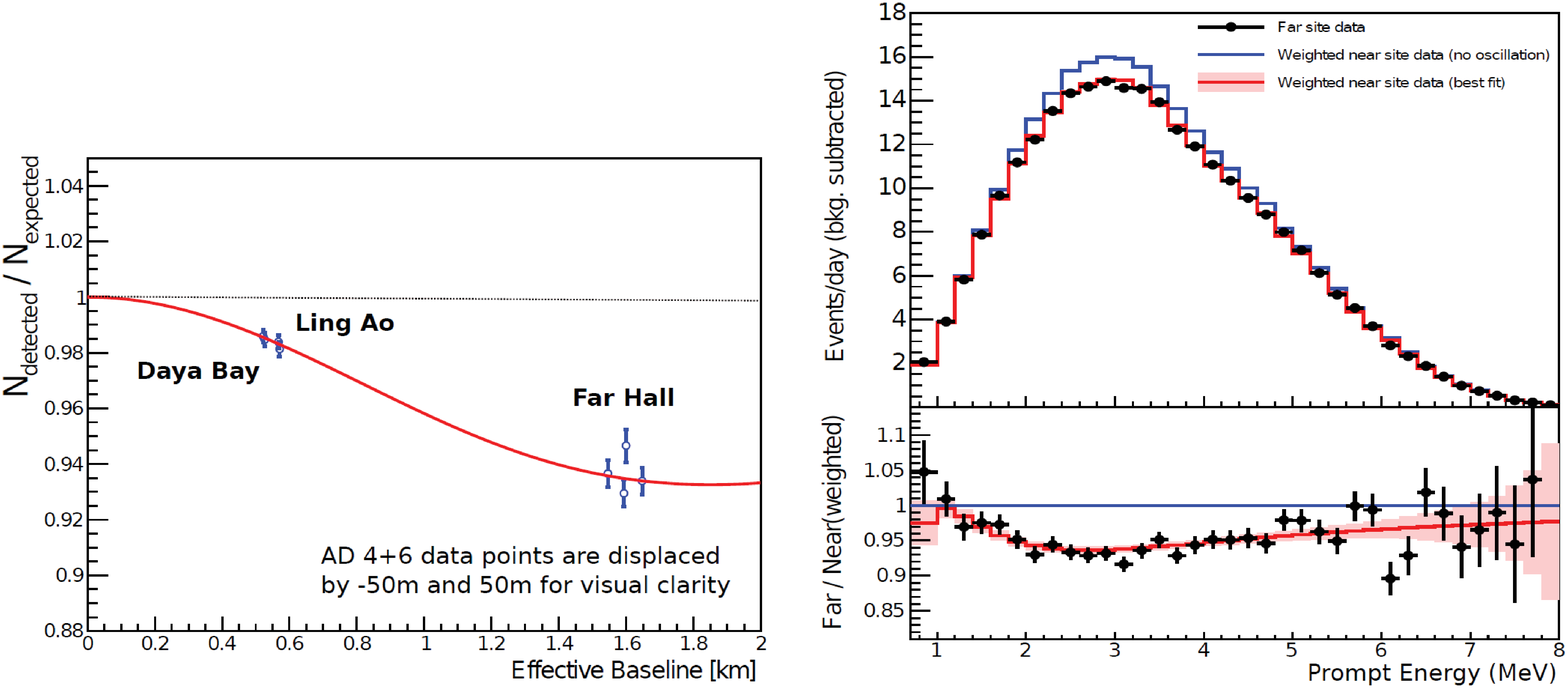}}
\caption{Daya Bay results reported in \cite{Neutrino14}. (left) Ratio of the detected to expected rates at the 8 antineutrino detectors (ADs) located in three experimental halls as a function of effective baseline. The expected signal accounts for the best-fit reactor antinuetrino flux normalization. The fitted
oscillation survival probability is given by the red curve. (right) The top panel shows the measured background-subtracted
spectrum at the far site compared to the expected spectrum based on the near site data both without oscillation and with
the best-fit oscillation included. The bottom panel shows the ratio of the far site spectrum to the weighted near site spectrum. The
red curve shows the expectation at the best-fit oscillation values from the rate and spectral analysis.}\label{Fig:DB1}
\end{figure}

The Daya Bay experiment took data with only 6 detectors  deployed from December 2011 to July 2012.  In summer 2012, two additional detectors were installed, one at the Ling Ao location and one at the far location, which completed the final 8 detector configuration of the experiment described above. Data taking resumed after October
2012. New results, based on the complete data set of the 6-AD period with the addition of the 8-AD period
from October 2012 to November 2013 (a total of 621 days) were recently reported \cite{Neutrino14}. The Daya Bay data display a substantial deficit in measured flux at the far site relative to the near sites, and also a distortion of the measured energy spectrum at the far site, consistent with the interpretation of neutrino oscillations as shown in Fig. \ref{Fig:DB1}.

The neutrino oscillation parameters are extracted from a fit to the rates and relative spectral shapes observed at the near and far sites, with the overall normalization of the flux as an independent parameter. The results yield the best fit values
\begin{eqnarray}
  \sin^2 2 \theta_{13} &=& 0.084 \pm 0.005 \\
  \Delta m_{ee}^2 &=& 2.44^{+0.10}_{-0.11} \times 10^{-3}~({\rm eV})^2
\end{eqnarray}
where $\Delta m_{ee}^2$ is defined by  $\sin^2 ( \Delta {m_{ee}}^2 L/ 4 E_\nu) \equiv \cos^2 \theta_{12} \sin^2 \Delta_{31} + \sin^2 \theta_{12} \sin^2 \Delta_{32}$.
This value of $\Delta m_{ee}^2$ is consistent, with comparable uncertainty, to the value of $\Delta m_{\mu \mu}^2$ determined by muon neutrino disappearance experiments.
\begin{figure}[ht]
\centerline{\includegraphics[width=0.8\textwidth]{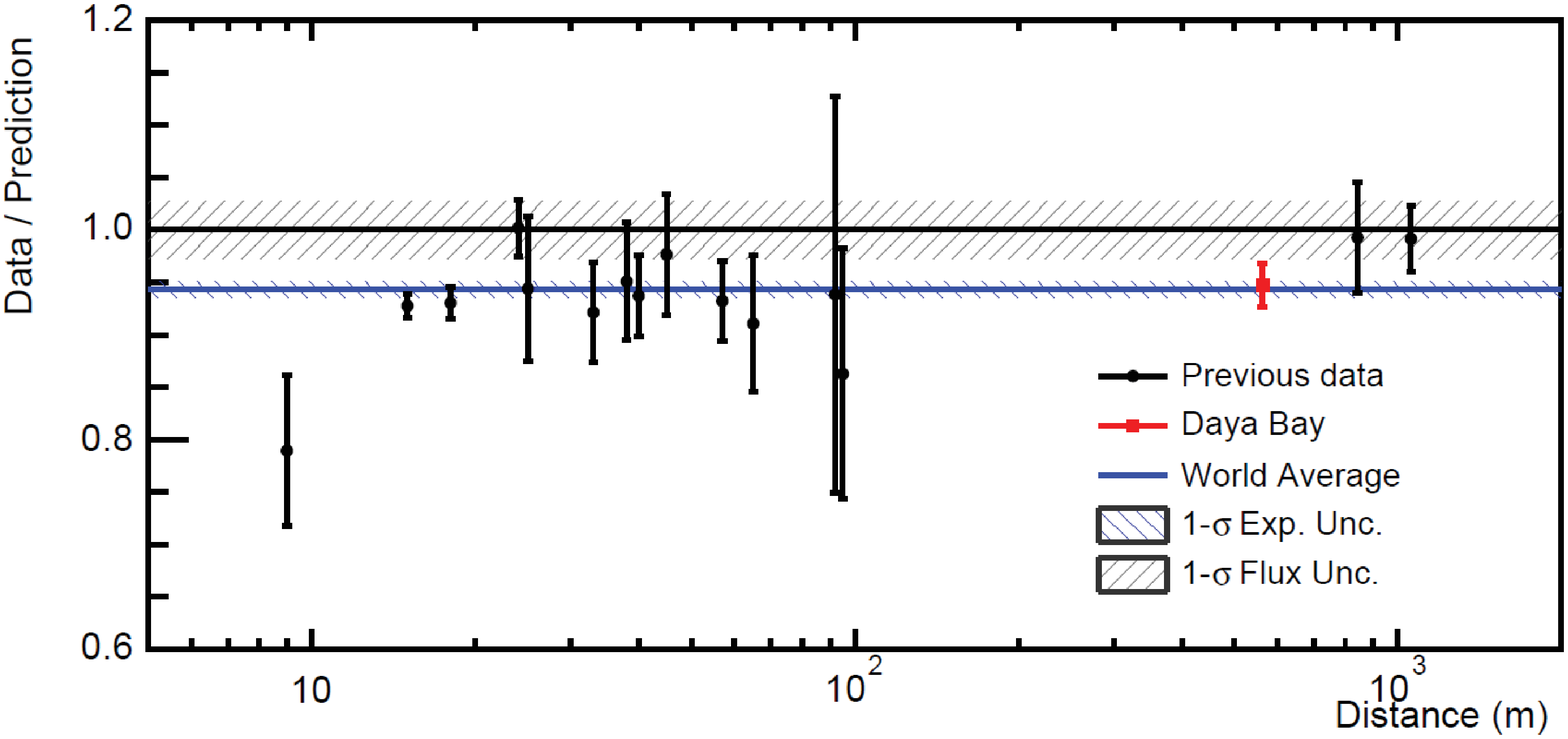}}
\caption{Measurements of antineutrino flux as reported in \cite{Neutrino14}. The reactor antineutrino interaction rate of the 21 previous short-baseline experiments as a function of the distance from the reactor, normalized to the Huber+Mueller model prediction \cite{Huber, Mueller}.  The Daya Bay result is placed at the effective baseline of 573~m. The rate is corrected for the  survival probability at the distance of each experiment, assuming standard three-neutrino oscillation. The horizontal bar (blue) represents the global average of all experiments and its 1$\sigma$ uncertainty. The 2.7\% reactor flux uncertainty is shown as a band around unity.}\label{Fig:DB2}
\end{figure}

The Daya Bay collaboration has also recently reported a measurement of the absolute flux of antineutrinos \cite{Neutrino14}, shown in Fig.~\ref{Fig:DB2}. This first precision measurement at larger average baseline (573~m) is consistent with 21 previous short baseline experiments, indicating a flux deficit of $5.3 \pm 2.2 \%$ relative to recent model predictions \cite{Huber, Mueller}.

New reactor neutrino projects \cite{JUNO}, JUNO and RENO50, are being planned by international collaborations to constrain neutrino oscillation parameters more precisely and to determine the mass hierarchy. The RENO50 experiment would be sited 50~km away from the Hanbit(Yonggwang)nuclear plant in South Korea. The JUNO experiment would be sited in southern China, 53~km equidistant from two new nuclear power plants currently under construction: Yangjiang (17.4~GWth) and Taishan (18.4 GWth). For JUNO, spherical 20kT liquid scintillator detector would be deployed at a depth of 700~m, with almost complete photocathode coverage to achieve the energy resolution of 3\% necessary to see the interference pattern in the energy spectrum for mass hierarchy determination. After 6 years of running, JUNO aims to achieve a $\Delta \chi^2 = 14$ determination of the mass hierarchy. In addition, the values of $\Delta {m_{12}}^2$, $\Delta {m_{23}}^2$ and $\sin^2 \theta_{12}$ will be measured with substantially higher precision than at present.

\begin{figure}[ht]
\centerline{\includegraphics[width=0.8\textwidth]{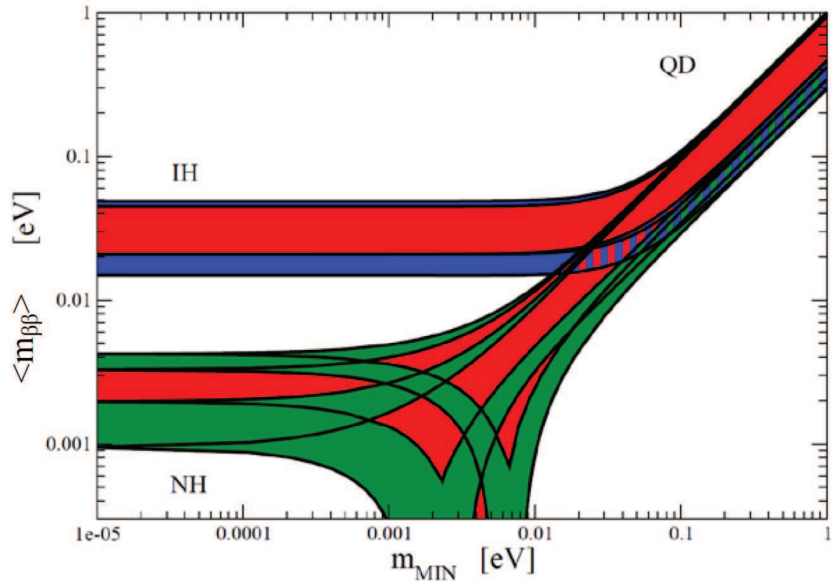}}
\caption{Allowed values of $\langle m_{\beta \beta} \rangle$ as a function of the  lightest  neutrino mass for the  inverted  (IH) and  normal (NH) hierarchies (QD stands for "quasidegenerate"). The red, blue and green bands correspond to different allowed regions for the unknown CP violating phases in Eq.~\ref{Eq:meff} and allowed 1$\sigma$ variation in the other known neutrino parameters. (From the Particle Data Group \cite{PDG}.) }\label{Fig:DBD1}
\end{figure}

\section{Neutrinoless Double Beta Decay}

Double beta decay is a rare transition between two nuclei with the
same mass number $A$ involving change of the nuclear charge $Z$
by two units. The decay can proceed only if the initial nucleus is
less bound than the final one, and both must be more bound than
the intermediate nucleus.
These conditions are fulfilled in nature
for many even-even nuclei, and only for them. Typically, the decay
can proceed from the ground state (spin and parity always $0^+$) of the
initial nucleus to the ground state (also  $0^+$) of the final
nucleus, although the decay into  excited states
($0^+$ or $2^+$) is in some cases also energetically possible.
Such nuclei can decay by the second order weak process, known as $2 \nu \beta \beta$ in which two antineutrinos as well as two electrons are emitted. The summed energy of the two electrons is a continuous distribution ranging from $2 m_e$ to the endpoint energy $E_0$ defined by the $Q$ value of the decay. This process conserves lepton number, takes place for both Dirac and Majorana neutrinos, and is the rarest decay process in nature for which half-lives have been measured.

For neutrinoless double beta decay, $0 \nu \beta \beta$, the distribution of summed $\beta$ energies would exhibit a distinctive monoenergetic peak at the endpoint $E_0$. If it occurs, this process implies nonconservation of lepton number and would imply that neutrinos were Majorana type fermions. The half life for this process can be written
\begin{equation}\label{Eq:NLDB1}
    { T_{1/2}^{0\nu} }^{-1}
~=~ G^{0\nu}(E_0,Z)
\left| M^{0\nu}  \right|^2
\langle m_{\beta\beta} \rangle^2
\end{equation}
where $G^{0\nu}$ is the exactly calculable phase space integral,
$\langle m_{\beta\beta} \rangle$ is the effective neutrino mass
and $M^{0\nu}$
is the nuclear matrix element (calculated using nuclear models).
The effective neutrino mass is
\begin{equation}
 \langle m_{\beta\beta} \rangle = | \sum_i {U_{ei}}^2 m_{\nu_i}  | ~,
\label{Eq:meff}
\end{equation}
where the sum is only over light neutrinos ($m_i < 10$ MeV), and contains the sensitivity to the neutrino masses and the elements of the neutrino mixing matrix $U_{ei}$. The $U_{ei}$ depend upon the mixing angles discussed above, but also two additional phases that do not contribute to neutrino oscillation experiments. The range of allowed values of $\langle m_{\beta\beta} \rangle$ is indicated in Fig~\ref{Fig:DBD1}.

As can be seen in Fig.~\ref{Fig:DBD1} the case of inverted mass hierarchy can lead to substantial values of $\langle m_{\beta \beta} \rangle$ even for very light values of the smallest neutrino mass. Thus there is considerable interest in performing experiments to address this region of parameter space. The current set of worldwide experimental efforts is summarized in Table~\ref{Tab:DBD}. These efforts aim to achieve a sensitivity exceeding $10^{26}$ years in the next 5 years, and provide crucial information on background reduction in order to assess the feasibility of scaling the next generation experiment up to the Tonne scale. Complete coverage of the inverted mass hierarchy band in Fig.~\ref{Fig:DBD1} will require multi-Tonne scale experiments.

\begin{table} [ht]
\begin{tabular}{|c|c|c|c|}
  \hline
   & & & \\
  Project & Isotope & Isotope fiducial  & Currently achieved  \\
   & & mass (kg) & $T_{1/2}$ limit ($10^{26}$ years) \\
    & & & \\
  \hline
   & & & \\
  CUORE & ${}^{130}$Te & 206 & $>0.028$ \cite{CUORE} \\
  Majorana & ${}^{76}$Ge & 24.7 &  \\
  GERDA & ${}^{76}$Ge & 18-20 & $>0.21$ \cite{GERDA} \\
  EXO200 & ${}^{136}$Xe & 79 & $>0.11$ \cite{EXO} \\
  NEXT-100 & ${}^{136}$Xe & 100 &  \\
  SuperNEMO & ${}^{82}$Se, + & 7 & $>0.001$ \cite{NEMO} \\
  KamLAND-Zen & ${}^{136}$Xe & 434 & $>0.19$ \cite{KZ} \\
  SNO+ & ${}^{130}$Te & 160 &  \\
  LUCIFER & ${}^{82}$Se & 8.9 &  \\
  \hline
\end{tabular}
\caption{Current double beta decay projects, the fiducial isotopic mass, and the currently achieved half-life limit (90\% CL). }
\label{Tab:DBD}
\end{table}

\section{Summary}

The study of neutrino properties with nuclear physics experiments is a very active field, with many experiments in progress and others in the planning stage. The absolute neutrino mass should be constrained by KATRIN to $0.2$~eV before the end of the decade. Beyond KATRIN, R\&D on the Project 8 method may offer a window to higher sensitivity measurements in the future. The present generation of reactor experiments will continue to reduce the uncertainties in $\theta_{13}$ and $\Delta {m_{ee}}^2$, and further study the flux and spectrum of reactor antineutrinos. A future experiment, JUNO, will be constructed in China with excellent potential to address the neutrino mass hierarchy. And an impressive suite of double beta decay experiments is underway that will extend the sensitivity towards the inverted mass hierarchy region in $\langle m_{\beta \beta} \rangle$.

These are indeed exciting times in the field of neutrino physics, with historic discoveries in the recent past, and the promise of much more to come in the future.

\section{Acknowledgments}

The recent Daya Bay results were first reported by Dr. Chao Zhang at the Neutrino 2014 conference, and I am very grateful to Dr. Zhang for supplying me with this material.
The financial support of the U.S. National Science Foundation grant PHY-1205411 is gratefully acknowledged along with the support of the U.S. Department of Energy, Office of Science, Office of Nuclear Physics under contract DE-AC05-06OR23177.

\pagebreak


\begin{footnotesize}



%

\end{footnotesize}


\end{document}